# Unconventional ferroelectricity in half-filling states of antiparallel stacking of twisted WSe$_2$

Liheng An[1,2#], Zishu Zhou[1,2#], Xuemeng Feng[1,2#], Meizhen Huang[1,2], Xiangbin Cai[1,2], Yong Chen[1,2], Pei Zhao[3], Xi Dai[1], Jingdi Zhang[1], Wang Yao[3], Junwei Liu[1,*], Ning Wang[1,2,*]

[1]*Department of Physics and Center for Quantum Materials, the Hong Kong University of Science and Technology, Hong Kong, China;*

[2]*William Mong Institute of Nano Science and Technology, the Hong Kong University of Science and Technology, Hong Kong, China;*

[3]*Department of Physics, University of Hong Kong, Hong Kong, China*

#Contributed equally to this work.

*Corresponding authors (emails: phwang@ust.hk (Ning Wang); liuj@ust.hk (Junwei Liu))

**Abstract:** We report on emergence of an abnormal electronic polarization in twisted double bilayer WSe$_2$ in antiparallel interface stacking geometry, where local centrosymmetry of atomic registries at the twist interface does not favor the spontaneous electronic polarizations as recently observed in the parallel interface stacking geometry. The unconventional ferroelectric behaviors probed by electronic transport measurement occur at half filling insulating states at 1.5 K and gradually disappear at about 40 K. Single band Hubbard model based on the triangular moiré lattice and the interlayer charge transfer controlled by insulating phase transition are proposed to interpret the formation of electronic polarization states near half filling in twisted WSe$_2$ devices. Our work highlights the prominent role of many-body electronic interaction in fostering novel quantum states in moiré-structured systems.

Stacking layered materials with small twist angles or with small lattice mismatches produce moiré superlattices that realize modulation of potentials at much larger spatial scale resulting in remarkable electronic properties [1, 2], for instance, correlated insulators [3], superconductivity states [4] and ferromagnetism [5, 6], etc. More recently, a new type of ferroelectricity has been created and identified through the stacking of layered van der Waals heterostructures of hexagonal boron nitride (BN) and semiconducting transition metal dichalcogenides (TMDCs) [7-13], of which the bulk structures forbids ferroelectricity. Ferroelectricity in twisted TMDCs is created by the structurally relaxed parallel stacking geometry and the moiré interface formed by the parallel stacking is also called 'R-stacking'. Two types of locally R-stacking domains coexist in one moiré unit cell, featuring opposite spontaneous polarizations of the same magnitude. The macroscopic ferroelectric polarization results from the dynamic bending of domain walls that microscopically favors polarization in one direction over the other [14].

According to the definition in previous publications [14-16], two identical 1H-WSe$_2$ layers stacked to each other at zero twisting angle will result in the parallel stacking interface. The antiparallel stacking interface is formed by rotating one of these two identical 1H-WSe$_2$ layers at 60° and stacking the two layers together. Opposite to parallel stacking geometry, in antiparallel stacking WSe$_2$, the electronic polarization locked to the local atomic registries is expected to be vanishingly small when local centrosymmetric geometry is fully respected [17]. Therefore, to date, neither theoretical nor experimental study has reported on the emergence of ferroelectricity in antiparallel stacked configuration. As we previously identified that strong electron-electron correlation effects occurred in twisted WSe$_2$ [15, 16, 18], it is anticipated that the correlation-driven insulating states in the moiré bands of antiparallel stacking WSe$_2$ may modulate interlayer charge transfer and even induce electronic polarization. It has been well acknowledged that the polarization in crystalline structures can be divided into ionic polarization and electronic state polarization [19]. In contrast to conventional displacive (ions or molecules) ferroelectricity, electronic polarization (the difference in polarization between two different states in the same structure) is rare in materials and normally occurs in a strongly correlated electronic system [20, 21].

In this study, we choose double bilayer WSe$_2$ to construct antiparallel stacking moiré superlattices and field-effect devices, of which the moiré flat bands with $\varGamma$ pocket holes contribute to the electrical transport. Stacking two identical bilayer WSe$_2$ with a twist angle near 0° results in the so-called antiparallel interface (Figure 1). In each moiré unit cell, there are three important high-symmetry stacking sites as labeled by AB, B$^{W/W}$ and B$^{Se/Se}$ separated by boundary regions (BR). AB sites are the energetically favorable regions (near 2H registry), whereas other regions correspond to higher energy states. The high symmetry sites periodically modulate the electronic states in real space and therefore produce moiré bands [22].

Atomically thin WSe$_2$ is mechanically exfoliated from high-quality bulk crystals of WSe$_2$. Field-effect devices are fabricated based on twisted double bilayer WSe$_2$ by using the tear-and-stack method [15, 18, 23]. To achieve a large size of uniform moiré lattices, we select a relatively large twist angle of about 4°. Atomically thin BN layers are used to form an encapsulated device structure. The electrical connection to the twisted WSe$_2$ channel is realized by Pt electrodes (with a matched work function to WSe$_2$) which offer a good efficiency for current injection. The device performance has been effectively improved by this kind of electrode design. The measured channel resistance is about 5 kΩ at a modest carrier density of $3\times10^{12}$ cm$^{-2}$ and the field-effect carrier mobility approaches 2000 cm$^2$V$^{-1}$s$^{-1}$. The device channel size is limited to $1\times10$ μm to achieve a good uniformity of the twist angle. The strong interlayer coupling between WSe$_2$ bilayers results in the rise of the $\varGamma$ valley band top (about 80 meV higher than that of the $K$ valley) [18]. Details of device fabrication and measurement principles are presented in Supplementary Information.

Our density functional calculation, in the absence of many-body Coulomb interaction, confirms the dependence of spontaneous electrical polarization on stacking registry, as found in the parallel stacking interface [14], but denies its existence in the antiparallel stacking interface of twisted double bilayer WSe$_2$ of arbitrary lattice match (Figure S2). Atomic-resolution electron microscopic imaging reveals several noteworthy features in our samples (Figure 2(a) and 2(c)). We first verify the twist angle and moiré periodicity by electron diffraction (Figure 2(b)). The general moiré structural features well match the model shown in Figure 1(a). However, we find that the AB regions (marked by hexagons) are expandable and the location of exact AB stacking (marked by the hexagonal dots and white arrows) is switchable. More analyses can be found in Figure S3. These structural instability features deviate from the theoretical model, and therefore hint on the necessity to incorporate new mechanisms for insights into the origin of the moiré band modulation near half filling under different electrical fields as discussed in more detail later.

The electronic states of the antiparallel stacking bilayer WSe$_2$ channel in the field-effect device shown in Figure 3 are tuned by the top ($V_{TG}$) and bottom ($V_{BG}$) electrical gates. $V_{TG}$ and $V_{BG}$ together can linearly tune the carrier density ($n = (C_{BG}V_{BG} + C_{TG}V_{TG})/e$) and displacement field ($\boldsymbol{D} = (C_{BG}V_{BG} - C_{TG}V_{TG})/2\varepsilon_0$) of this p-type semiconductor electronic system, where C and V represent device capacitances and gate voltages, $e$ is the elementary charge and $\varepsilon_0$ is the vacuum permittivity. A negative $V_{TG}$ (-12 V to -14 V) is first applied in order to achieve good electrical contacts to the twisted WSe$_2$ and injection of holes in the device channel. Forward scanning of $V_{BG}$ (from -60 V to +60 V) is to release holes from the channel and increase $\boldsymbol{D}$. At $V_{BG}$ = +60 V, the hole density in the channel is very low. The typical electrical transport characteristics of the p-type antiparallel twisted WSe$_2$ devices measured by a four-probe configuration at 1.5 K are shown in Figure 3, in which a quick decrease of the resistance ($R_{xx}$) by decreasing $V_{BG}$ from +60 V indicates that the Fermi level touches the edge of the topmost moiré band. By further decreasing $V_{BG}$, two metallic and two insulating states are detected. The metallic and insulating states are verified by measuring their $R_{xx}$ at different temperatures as demonstrated in Figure 4(a). We further verify that the two insulating states are from the correlation-induced splitting of the topmost moiré band [2, 23, 24], corresponding to the half and full filling states respectively. The full filling gap is resolved at a carrier density of $n = 9.8 \times 10^{12}$ cm$^{-2}$ with two holes per unit cell to fully occupy the first moiré band [25]. The resistance peak emerging near $n =$ 5~6$\times 10^{12}$ cm$^{-2}$ represents the important half filling insulating states driven by strong correlation effects. By forward-and-backward scanning $V_{BG}$ for several times, an obvious resistance hysteresis is repeatedly observed around half filling states (Figure 3). At half filling, when $V_{TG}$ is set at -13.8 V, the forward scanning (-60 V to +60 V) leads to a high resistance state of about 12.5 kΩ, while the backward scanning yields a low resistance state of about 10 kΩ. Such a resistance difference can be considered as the appearance of internal electric fields [13] from charge polarization (perpendicular to the moiré lattice plane) states in the system. We have carried out each measurement for five times to ensure the reproducibility of the hysteresis loops. In addition, measurements of the I-V curves at the half filling states also confirm that the forward and backward scanning of gate voltages result in different channel resistances (Figure S9). To rule out the possibility that the BN dielectric layers or metal lead interfaces in the devices could potentially generate a similar hysteretic resistivity, we fabricated and measured reference devices for comparison. We also performed different gate scanning rates during hysteretic resistivity measurements and did not observe any impurity/charge trapping effect. All these additional experiments evidence that BN or charge trapping effects do not play a role in the observed hysteretic resistivity near half filling states in our samples (See details in the Supporting Information).

The appearance of the electrical polarization characteristics in the antiparallel stacking interface structures of WSe$_2$ is abnormal. Different from conventional [19] and newly discovered layered van der Waals ferroelectricity [7-13, 26-29], the electrical polarization in our samples is highly relevant to correlation effects which are temperature dependent. As shown in Figure 4(a), the half filling insulating states gradually disappear when temperature is higher than about 40 K. The largest hysteresis we observed at 1.5 K also gradually disappears when temperature approaches to about 40 K (Figure 4(c)). All these phenomena are very different from that of the ferroelectricity observed in the parallel stacking geometry of TMDCs where the spontaneous electronic polarization survives at room temperature, not relevant to filling states at all, further differentiating the origins of the ferroelectricity effects between the two twist moiré systems.

Here, we try to elucidate the abnormal electrical polarization characteristics based on the following facts. The transport data are from the contribution of the $\Gamma$ valley edge of $WSe_2$ valance bands [18] and the ultra-flat moiré bands (survived up to 4° twist angles [30]) are spatially associated with the AB sites in the antiparallel stacking interface of $WSe_2$ [16]. AB sites form a triangular lattice which can be considered as a mimic triangular lattice of the Hubbard model (Figure 5(a)). The topmost moiré band is separated from the rest and its charge distribution is tightly localized at the moiré potential minima of AB sites. Double occupancy of AB sites is suppressed due to the strong on-site Coulomb repulsion potential $U$, resulting in moiré band splitting to lower (LHB) and upper (UHB) Hubbard bands [31].

Ideally, the carrier densities at full/half filling of the Hubbard bands are purely determined by the moiré periodicity and obtained by $n_0 = 4/(\sqrt{3}\lambda^2)$, where $\lambda$ is the moiré superlattice constant. We first focus on the $R_{xx}$ peak positions of full filling states as a reference to test the functionality of the two gates. By setting $V_{TG}$ and scanning $V_{BG}$ (Figure 4(b)), we confirm that the injected density $n$ (hole carriers in p-type $WSe_2$) at full filling (the gap position) as characterized by $R_{xx}$ peaks are all correctly tuned by the two gates. This is reflected by the change of the full filling peak positions as outlined by the inclined straight dashed line in Figure 4(b). By increasing $V_{TG}$, the full filling peak positions shift to right side, meaning that along this dashed line, the carrier density is unchanged. Because both full and half filling densities are fixed by the moiré lattice geometry [32], the $R_{xx}$ peaks at half filling should also follow a straight line parallel to the full filling peak line. However, the peak positions at half filling obviously shift to left side (towards to higher hole concentration regions). This suggests that when increasing the $D$ field, more holes are needed to be injected into the system in order to completely fill the LHB as displayed more clearly in Figure 5(c). In addition, we noticed that the $R_{xx}$ at half filling decreases by increasing $D$, which could be interpreted as that the Hubbard gap $U$ becomes narrow accordingly. These interesting experimental data suggest following two possible physical pictures: (1) $D$ enhances new electronic states involved in the half filling states in the moiré system; or (2) increasing $D$ could narrow the Hubbard band gap $U$.

We now provide a theoretical picture to intuitively interpret the correlation-effect-induced electronic polarization based on the interlayer charge transfer controlled by electric fields and insulating phase transition. For a large $D$ field applied to the double bilayer twisted device, it is possible that the entire low-energy moiré bands become layer polarized, as reported in the unconventional electronic polarization in bilayer graphene heterostructures [9]. Similarly, electrons occupying a moiré band at low energy could locate on a specific $WSe_2$ layer in the device. It is also reasonable to assume that LHB/UHB locates at the interface between the two inner $WSe_2$ layers (Figure 3). In our device design, the electrodes directly connect to the bottom $WSe_2$ layer and holes are firstly filled preferably in the bottom layer. We then propose that the observed $D$-dependent shift of the half filling peak is due to the involvement of new electronic states. Figure 5(b) schematically shows the additional band (B) coexisted with the Hubbard bands which contributes to the new electronic states. This band can be considered as the contribution from the bottom $WSe_2$ layer directly connected to the electrodes for hole injection. The inner two layers which host LHB/UHB states are actually not directly connected by the electrodes.

At a fixed $D$ field, the backward scanning of $V_{BG}$ (starting from +60 V) means filling holes into the electron-occupied LHB first (Figure 5(b)). In our p-type $WSe_2$ moiré system, the charge carriers which contribute to the transport measurement are holes in the valance band. Before adding holes into the valance band, all states are occupied by electrons. We use solid color (black and yellow) to illustrate the electron occupied states in the LHB and UHB in Figure 5(b). By turning the Fermi level, holes are firstly injected into the LHB and then into the UHB. Because of the involvement of the B-band, the apparent Hubbard gap position may shift to left side. During backward $V_{BG}$ scanning, hole carriers transfer from the B-band to LHB. This interlayer charge transfer is driven by the potential associated with the gates and $D$ field [33]. The bottom $WSe_2$ layer offers the B-band which is p-type under the same gating condition (larger than the threshold gate voltage). Figure 5(b) shows the band alignment and the carrier transfer direction between the bands. The hole filling process allows current to flow easily from B-band (p-type) to the Hubbard bands (p-type) across the spatial layer interface (similar to a p-p isotype junction [34]). Near half filling, however, the states at the bottom edge of LHB are electron-like (n-type). Holes transfer across a p-n anisotype junction (the forward biasing of a p-n junction) is less resistive. Therefore, filling LHB/UHB driven by backward $V_{BG}$ scanning is less resistive. For the forward $V_{BG}$ scanning, however, holes are released from LHB/UHB to the B-band. Near half filling, the p-n anisotype characteristic at the interface severely restricts holes reversely flowing through the anisotype junction (analogous to the reverse biasing of a p-n junction). The consequence is that more holes are retained in the moiré layer when the forward scanning is near half filling, while the holes in the B-layer are always released to the electrodes normally during $V_{BG}$ scanning. Therefore, the charge polarization between the moiré layer and the B-layer is established. The extra holes in the moiré layer countervail only a part of $D$ field strength (Figure S11(b)), resulting in a higher $R_{xx}$ compared to that in the opposite direction of backward scanning $V_{BG}$. This is because $R_{xx}$ is inversely proportional to the strength of the $D$ field as revealed experimentally in Figure 5(c). This mechanism supports the transport data of the $R_{xx}$ hysteresis in Figure 3. We now conclude that the two electrical polarization states are the result of more versus less occupation of the Hubbard bands driven by the interlayer carrier transfer modulated by the insulating phase transition during $D$ field scanning near half filling states. This abnormal electronic polarization effect has been repeatedly observed in the devices with twist angles ranging from 3.8-4.2º (see more data in the Supporting Information).

Electronic ferroelectricity caused by electron correlation has been proposed and studied previously based on Hubbard models in different bulk material systems [20, 21], in which parameters of Coulomb repulsion potential $U$ and hopping energy $t$ are critical. In our moiré system, the expandable AB sites driven by correlation effects or $D$ fields [26] could modulate $U$ and $t$. This is because the local AB sites (near 2H registry) in our samples is expandable (also deviating from an ideal local 2H registry) involving the atomic rearrangement of the boundaries to surrounding regions. Therefore, the expanded local 2H registry together with the boundaries surrounding these regions do not hold the centrosymmetric geometry according to our electron microscopy results (see more details in the Supporting Information). The expansion of AB sites (increasing the energy favorable area) driven by the $D$ field could provide the driving force for decreasing $U$, since increasing AB site area can effectively lower $U$ to host two electrons with opposite spins in one AB site. At present, it is still not clear whether the deviation of the half filling $R_{xx}$ peak under different $D$ fields is solely due to the interplay between the Hubbard bands and B-band or partially due to the $D$ field modulation (e.g., $D$-field-induced electronic state instability) of the electronic states in the Hubbard bands. To this end, further experimental and theoretical investigations are needed. Our discovery of the abnormal ferroelectricity behaviors suggests a new platform for further exploring the flat band properties tunable by an electric field, particularly the correlation-induced insulating states at the half filling states adjacent to the superconductivity states in this two-dimensional moiré system.

## Acknowledgements

We are grateful to the technical support from Dr. Yuan Cai from the Materials Characterization and Preparation Facility at the Hong Kong University of Science and technology.

## Funding

This work was supported by grants from the National Key R&D Program of China (2020YFA 0309600), the Hong Kong Research Grants Council (Project Nos. AoE/P-701/20, C6025-19G, 16305919 ECS26302118, 16303720, 16305019, 16306220 and N_HKUST626/18), National Natural Science Foundation of China (NSFC20SC07) and the William Mong Institute of Nano Science and Technology.

## Author contributions

N.W. initiated and supervised the project. L.A. designed the research, device structure, measurement and data analysis. L.A., Z.Z., X.F. fabricated device and collected data. L.A. and N.W. wrote the manuscript. M.H., J.Z., X.D. and J.L. contributed to data analysis. P.Z. and W.Y. contributed to calculation. X.C. and Y.C. contributed to sample preparation.

## Conflict of interest

The authors declare that they have no conflict of interest.

## Data availability

The original data are available from corresponding authors upon reasonable request. Details of device fabrication, measurement principles and proposed mechanisms are presented in **Supplementary Information**.

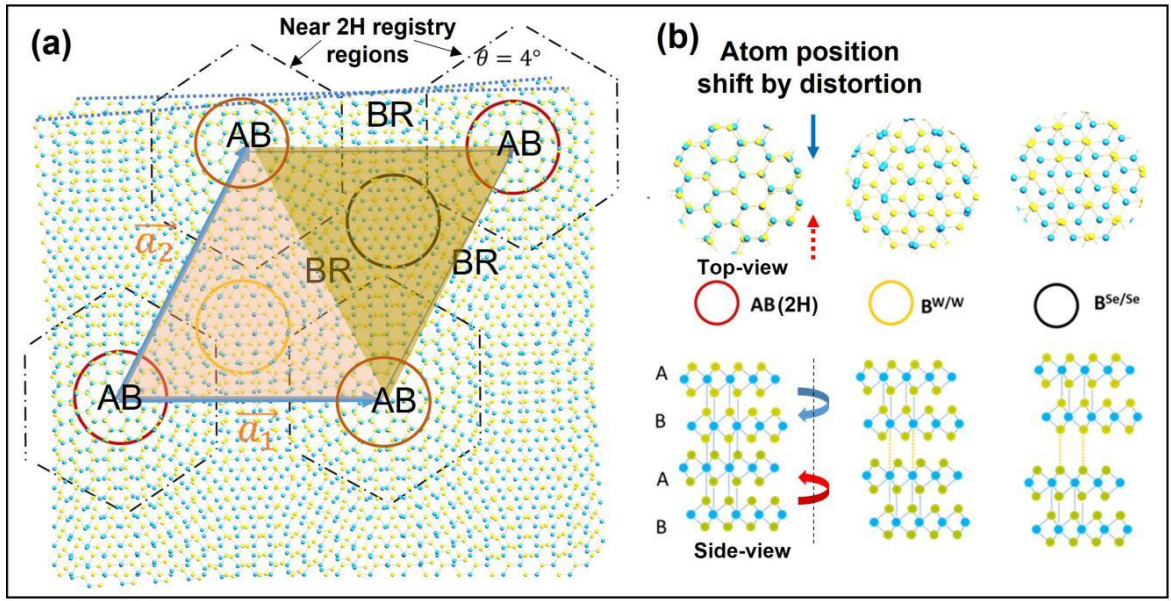

**Figure 1** (a) Moiré superlattice structure constructed by antiparallel stacking of twisted double bilayer WSe$_2$. The circles denote the high symmetry stacking sites. The hexagons (dashed lines) indicate the near 2H registry regions. BR indicates the boundary regions. (b) The atomic configurations of the high symmetry stacking sites of AB, B$^{W/W}$ and B$^{Se/Se}$ and their side views of these sites. Blue and red arrows denote the atom positions shifts from top and bottom layers of WSe$_2$.
8

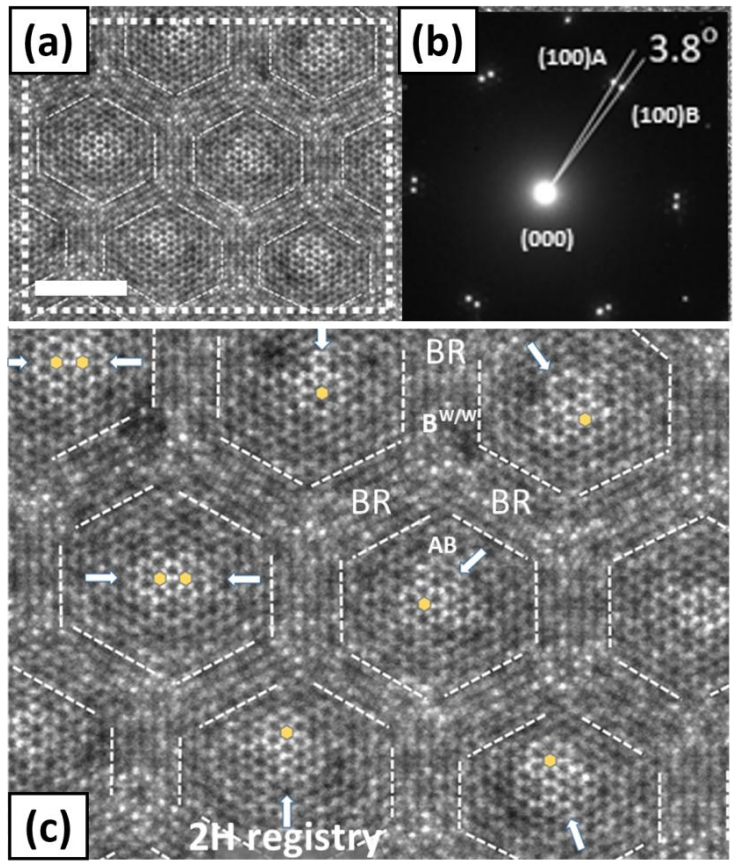

**Figure 2** (a) Transmission electron microscopy (TEM) image of the antiparallel twisted double bilayer WSe$_2$. (b) Electron diffraction pattern taken from the area in (a). The twist angle is determined to be about 3.8°. (c) An enlarge image of the region marked by the white dashed rectangle in (a). The positions of exact 2H stacking are marked by yellow hexagons (dashed lines). The white thick arrows indicate the directions of the 2H registry shifts. The scale bar is 3 nm.

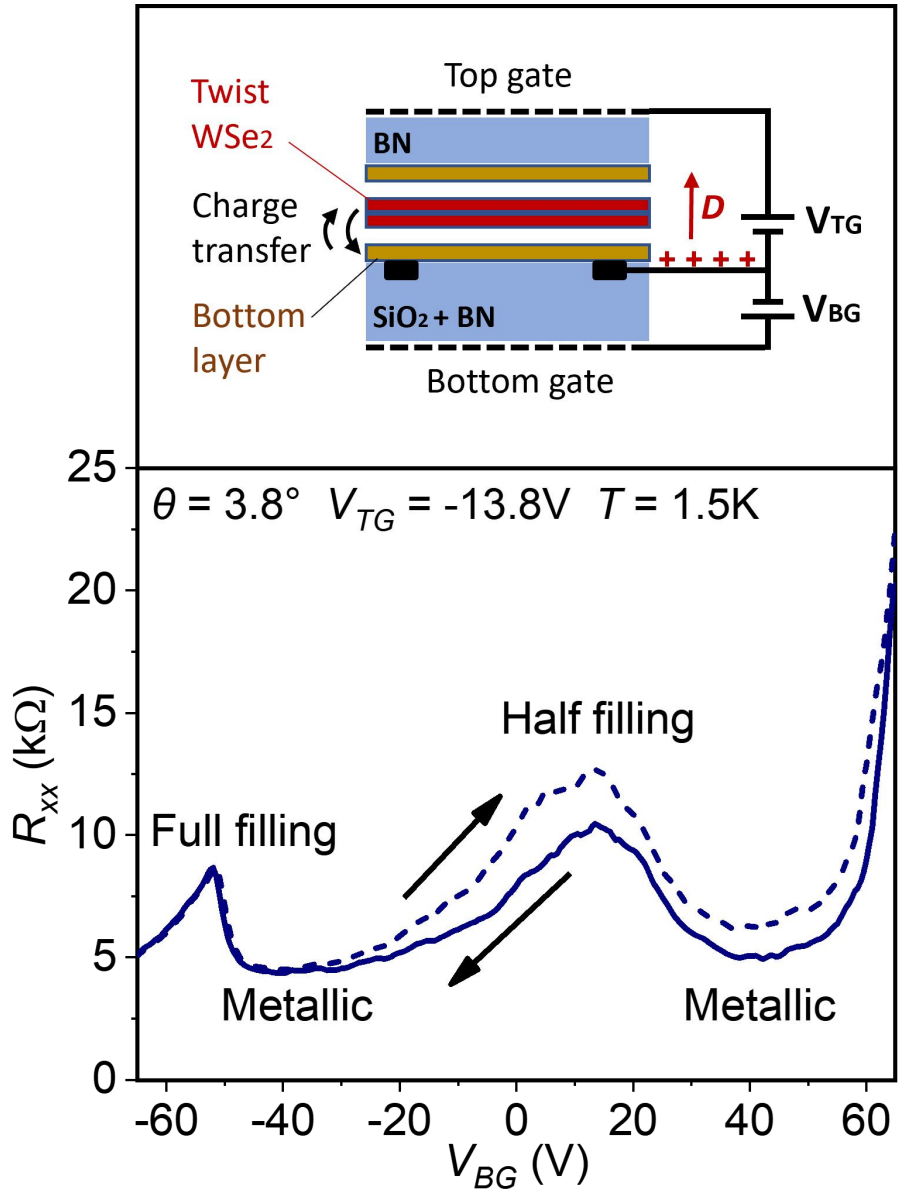

**Figure 3** The structure of the field-effect device built based on antiparallel stacking of twisted double bilayer WSe$_2$ and the typical resistance hysteresis measured in the device. The device is double gated by the top and bottom gates in order to tune the carrier concentration and the displacement field (***D***). By fixing the top gate (V$_{TG}$) at -13.8 V, scanning the bottom gate (V$_{BG}$) results in the change of the filling states. Forward-backward scans (indicated by the dashed/solid lines) of V$_{BG}$ show a clear resistance hysteresis.

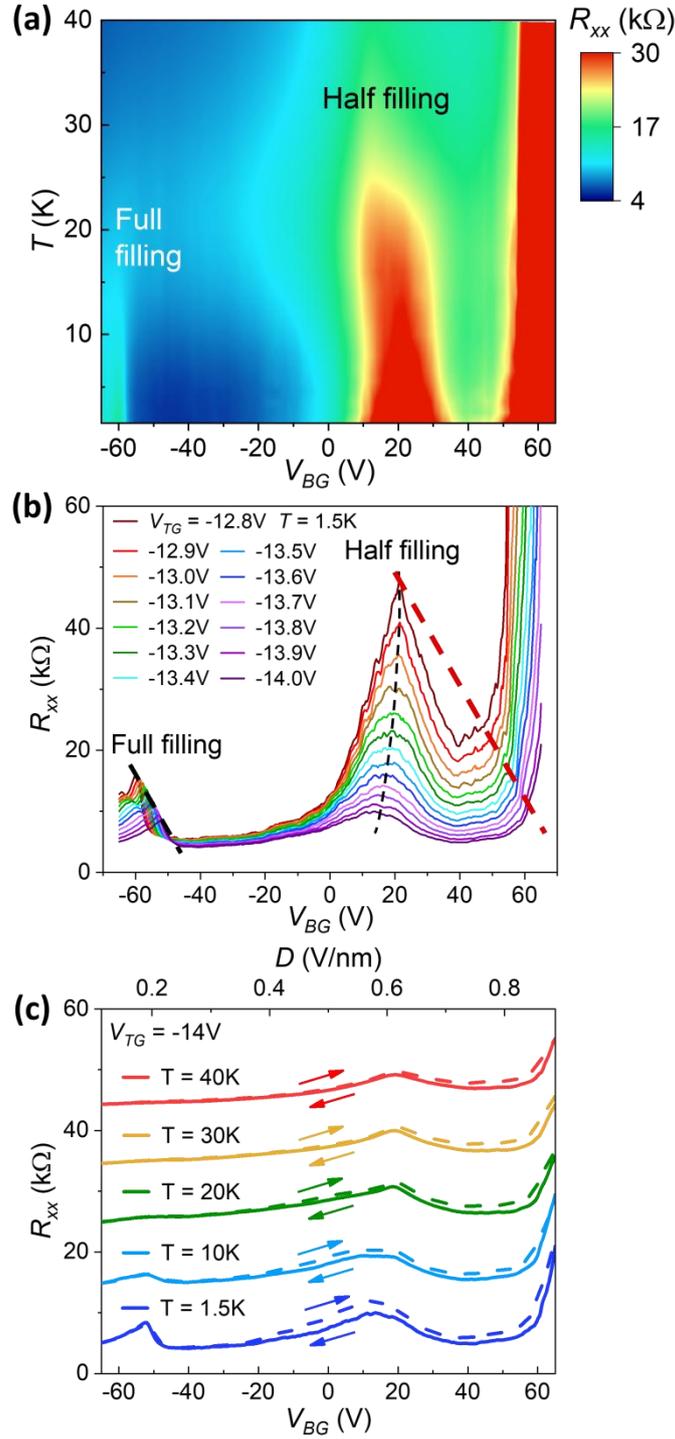

**Figure 4** (a) Phase diagram plotted based on transport measurement. The half filling states gradually disappear when temperature is higher than 40 K. (b) Displacement field effects (by setting $V_{TG}$ at different voltages and scanning $V_{BG}$) on the full filling and half filling states. The full and half filling states behave differently by changing the displacement fields. The black/red thick dashed lines indicate the ideal full/half filling peak positions. The black thin dashed line indicate the peak positions of half filling states measured. (c) Temperature dependence of the resistance hysteresis. The hysteresis loops gradually disappear at about 40 K.

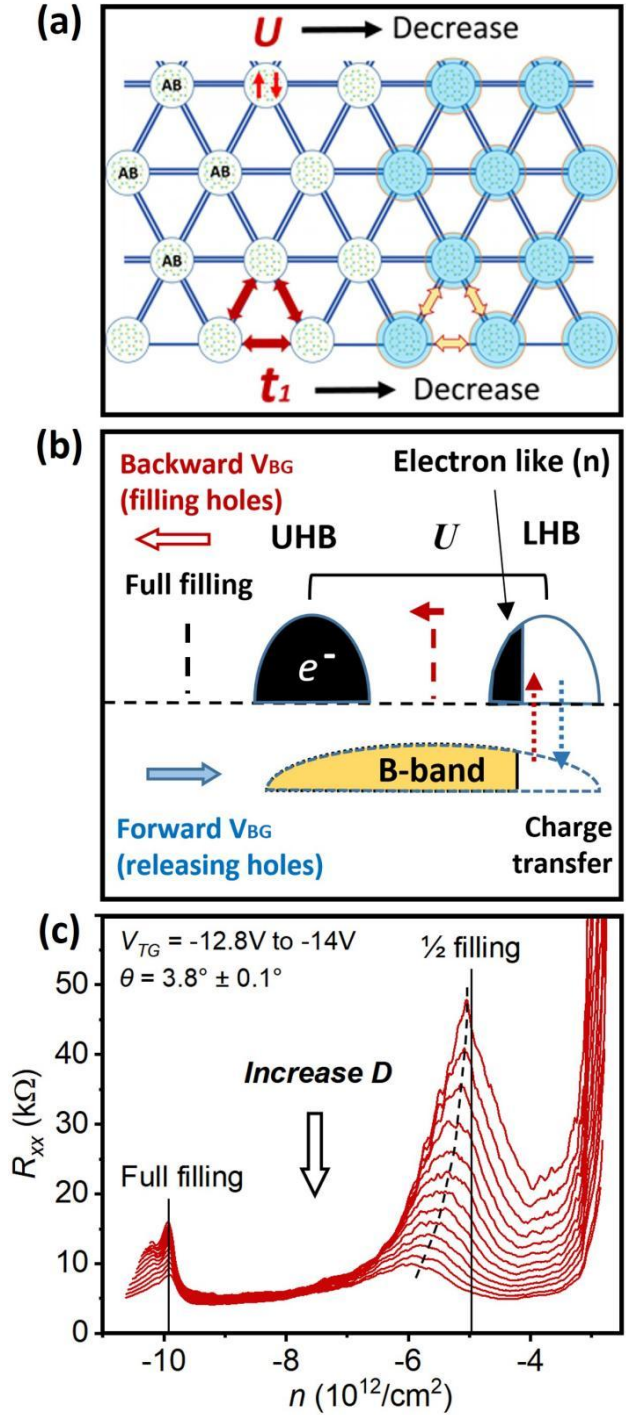

**Figure 5** (a) A mimic triangular lattice of the Hubbard model for the AB sites and the variation of $U$ and $t$ due to AB site expansion. (b) Carrier transfer between the B-band and LHB/UHB. Forward scanning of $V_{BG}$ results in the releasing of holes and the backward scanning of $V_{BG}$ results in the filling of holes in the bands. Charge transferring between the B-band and LHB/UHB shows different behaviors near the gap region of the Hubbard bands. (c) Displacement field effects on the full filling and half filling gap positions. The full filling states are independent of the ***D*** field as indicated by the thin solid line. However, the half filling is ***D***-dependent which deviates from the ideal positions as indicated by the right side thin solid line.

# Supplementary Information

## Unconventional ferroelectricity in half filling states of antiparallel stacking of twisted WSe$_2$

Liheng An[1,2#], Zishu Zhou[1,2#], Xuemeng Feng[1,2#], Meizhen Huang[1,2],

Xiangbin Cai[1,2], Yong Chen[1,2], Pei Zhao[3], Xi Dai[1], Jingdi Zhang[1], Wang Yao[3],

Junwei Liu[1, *], Ning Wang[1,2,*]

([#] These authors contribute equally to this work)
*Corresponding authors (emails: phwang@ust.hk (Ning Wang); liuj@ust.hk (Junwei Liu))

[1]Department of Physics and Center for Quantum Materials, The Hong Kong University of Science and Technology, Hong Kong, China.

[2] William Mong Institute of Nano Science and Technology, The Hong Kong University of Science and Technology, Hong Kong, P.R. China.

[3]Department of Physics, University of Hong Kong, Hong Kong, China.

### A. Device fabrication

Atomically thin WSe$_2$ is mechanically exfoliated from bulk crystals (from 2D Semiconductors). The bilayer WSe$_2$ is identified by optical microscopy and photoluminescence techniques. The bottom hexagonal boron nitride (hBN) is around 25 nm thick. The bottom hBN is pre-patterned and selectively etched down to the SiO$_2$ by plasma treatment in CHF$_3$. Then, Cr/Pt 10 nm/20 nm is deposited to form the bottom electrodes. We fabricated the twisted double bilayer WSe$_2$ devices by using the tear and stack method as previous work introduced [S1-3]. Then polypropylene carbonate is used to tear one part of the bilayer WSe$_2$. The bottom flake is rotated by a small angle and the top flake is used to pick up the bottom flake. The whole stacked structure is placed on the pre-patterned bottom electrodes to form the contacts to the twist structure (FIG. S1). Finally, another flake of hBN (30 nm – 50 nm) is transferred onto the top surface of twist WSe$_2$, and a thin layer of Cr/Au (10/70 nm) is deposited on the top surface of hBN to form the top gate. Pt layers directly deposited onto the top surface of the bottom hBN generally result in gas trapped around the bottom electrodes on the sample interfaces. The trapped gas, after performing a thermal annealing, accumulates at the electrode edges, leading to detach of the metal and the sample interface. The large real space gaps between the metal and the sample is the main reason for generating a huge contact resistance and thus degrade the device quality. Here, by deep etching of the bottom hBN and depositing Pt electrodes, we improved the evenness of the bottom Pt electrodes, hBN bottom surfaces and the current injection efficiency from the metal electrodes into the atomically thin WSe$_2$. For electron microscopy investigation, the twisted structure was dropped onto a holey carbon grid, cleaned by different chemical solutions, and then dried in vacuum.

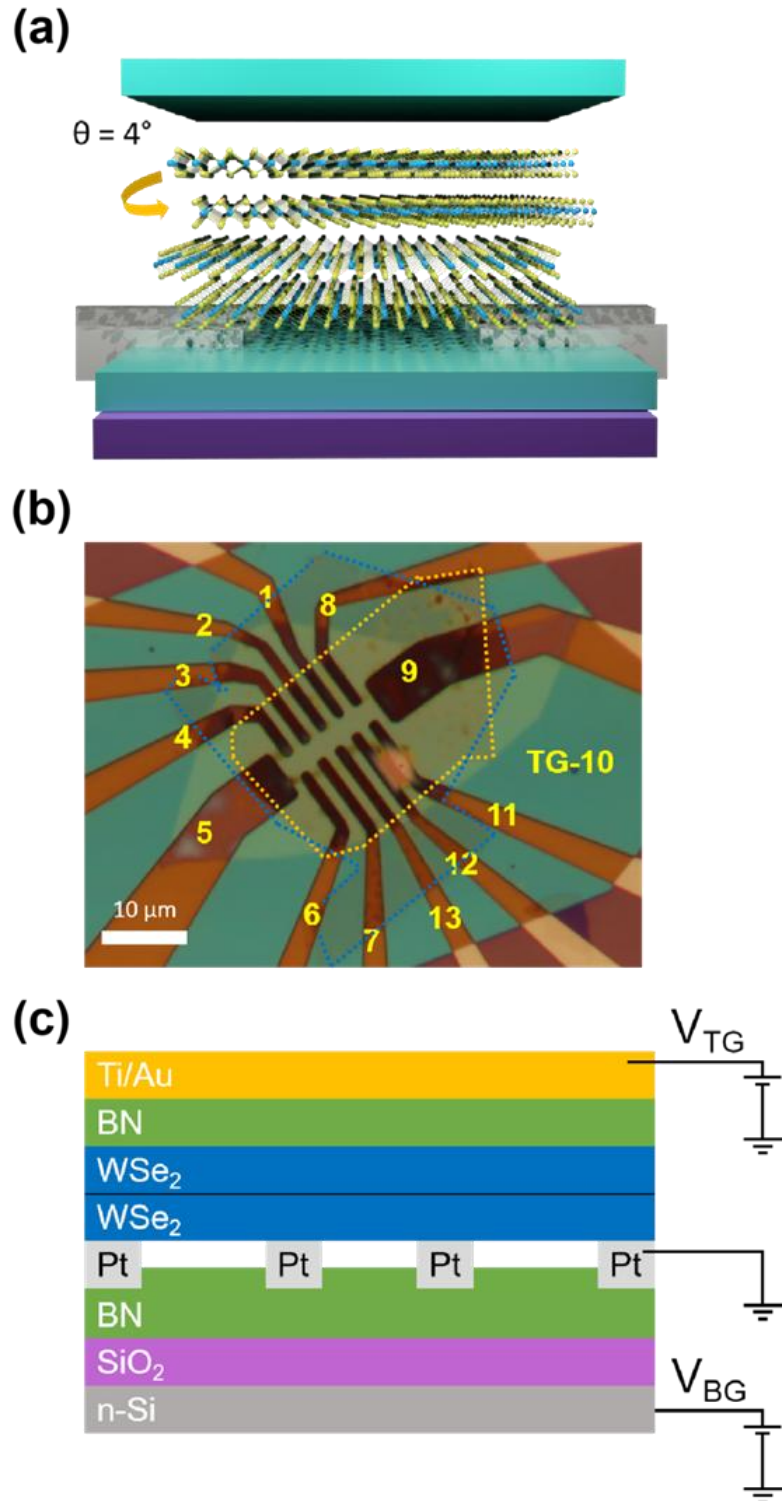

FIG. S1. (a) Device structure of antiparallel stacking of bilayer WSe$_2$. (b) Twist WSe$_2$ structure placed on the metal electrodes before covering the hBN and metal gate. (c) Cross-sectional view of the twist WSe$_2$ device design.

## B. Improvement of device performance

To ensure the quality of the devices, two layers of hBN are used to fabricate the fully encapsulated structure (FIG. S1(c)). The misalignment of the top and bottom hBN is carefully considered to avoid the formation of any small angle twisted hBN/WSe$_2$ interfaces. In fact, the transport characteristics measured in the devices are dominated by the moiré interface properties especially the strong half filling ($\nu=1$, one hole per moiré unit cell) insulating states [S4], thus the hBN interface effects are ignorable. We use the bottom-electrode design to make the contacts to the channel of the bottom twist layer (See FIG. S1). We use the top and bottom gates to control the out-of-plane DC electric field applied to the moiré channel and change the carrier concentration. The device performance has been effectively improved by this kind of device design. The measured channel resistance is about 5 kΩ at a modest carrier density of $3\times10^{12}$ cm$^{-2}$ and the field-effect carrier mobility approaches 2000 cm$^{-2}$V$^{-1}$s$^{-1}$. The device channel size is limited to $1\times10$ μm to achieve a good uniformity of the twist angle. Such small-sized samples integrated with bottom and top metal electric gates therefore limit the detection of the ferroelectric property at cryogenic temperatures by conventional techniques. Our transport data are from the $\Gamma$ valley. The strong interlayer coupling between WSe$_2$ bilayers results in the rise of the $\Gamma$ valley band top (about 80meV higher than that of the $K$ valley) [S3].

### C. Computation results

Based on the same theoretical model presented in Ref. [S8], we examined the difference vacuum levels $\Delta E_{vac}$ on the two sides of the 2L+2L WSe$_2$ with lattice-matched antiparallel stacking. We find that $\Delta E_{vac}$ is 0 (cf. FIG. S2), indicating the electrical polarization is forbidden. Notably, the electrical polarization at the interface of 2L+2L WSe$_2$ has invisible dependence on each stacking registry $r_t$. In this calculation, interaction effects have not been considered.

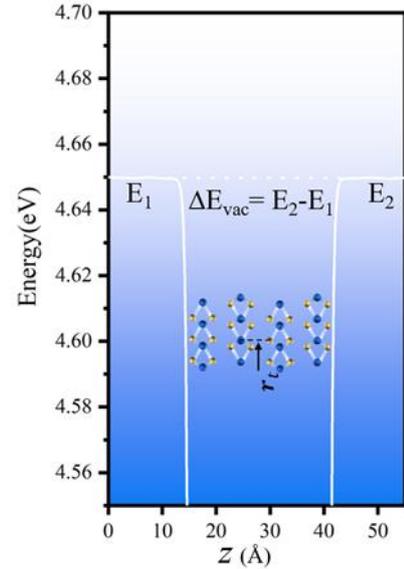

FIG. S2. *The difference in the vacuum levels on the two sides of the 2L+2L WSe$_2$, $r_t$ indicates the stacking registry of double bilayer WSe$_2$.*

We have carried out measurements in different samples with antiparallel 2L+2L stacking geometry (FIG. S2) at twist angles ranging from 3.8-4.2 degrees. These samples did not show hysteretic resistivity at room temperature since the half filling states did not form at room temperature. While at cryogenic temperatures, half filling sates emerge in these samples accompanied with the hysteretic resistivity effects. Theoretically, this ideal antiparallel 2L+2L stacking geometry with a local 2H registry does not have out-of-plane polarization. Therefore, in principle, the gate voltage should not cause expansion of the local 2H registry. This also implies that "ideal" half filling states should be displacement field independent, and there should not be electrical polarization/ferroelectricity effect in this system. However, interaction effects may induce symmetry breaking and generate electronic polarization.

### D. Atomic structure of 2L+2L twisted WSe$_2$

We identified by high-resolution electron microscopy that the local AB sites (near 2H registry) in our samples is non-uniformly expandable (also deviating from an ideal local 2H registry) involving the atomic rearrangement of the boundaries to surrounding regions. Therefore, the expanded local 2H registry together with the boundaries surrounding these regions do not hold the centrosymmetric geometry according to our microscopy results. On the other hand, it is reasonable to assume that symmetry breaking (or atom position deviation) effects could generate charge transfer between layers and induce electronic polarization. As demonstrated by electron microscopy, the expandable 2H sites already break the local 2H centrosymmetric geometry.

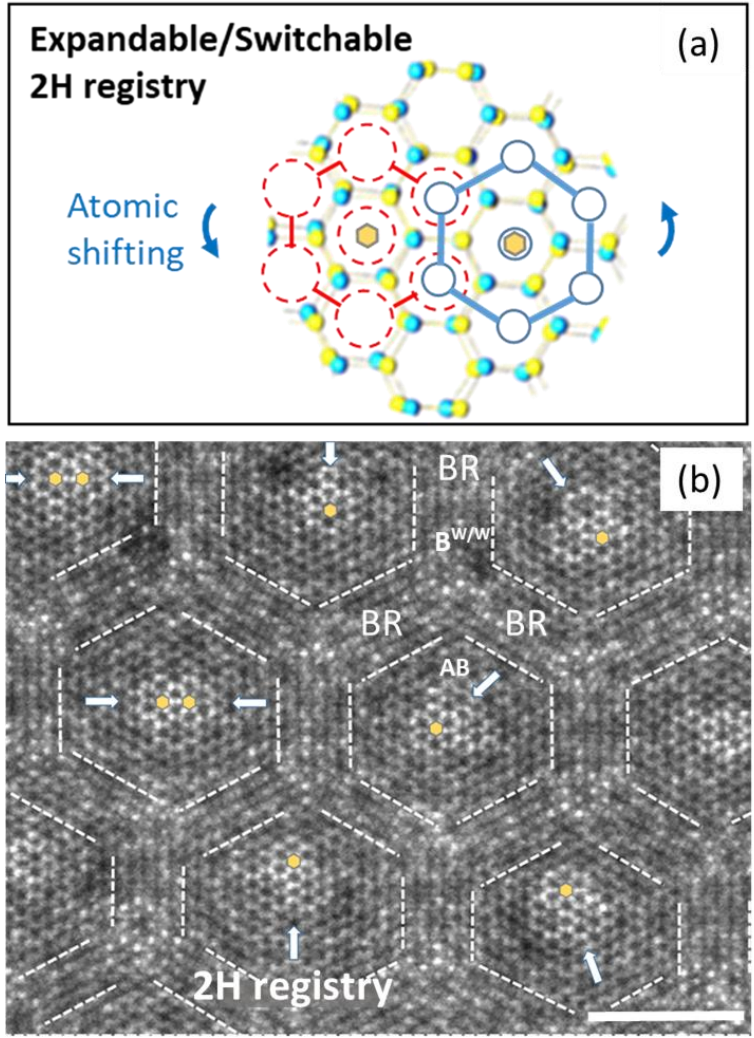

FIG. S3. (a) Schematic illustration of the expandable and switchable 2H registry induced by atomic position shifting. (b) An enlarge TEM image showing that the exact 2H registry positions as marked by yellow hexagons. The white thick arrows indicate the expansion or shift of the 2H registry is non-uniform. The scale bars are 3 nm.

By analyzing the atom positions in transmission electron microscopy (TEM) images, we identify the details of the 2H registry sites (the AB regions marked by hexagons). An ideal 2H registry site is displayed as one "hexagonal-flower-like" pattern in the atomic image with strong and sharp contrast (centered at a small yellow hexagon), in which the strong bright dots indicate highly out-of-plane aligned atoms. The larger the area covered by the sharp hexagonal-flowers, the more expended area occupied by the 2H registry. Obviously, many 2H sites show two or three strong and sharp contrast of the hexagonal-flowers emerging together due to the expansion of the 2H registry area. FIG. S3(a) illustrates the mechanism of the expandable or switchable 2H registry which is useful for understanding the modulation of the Hubbard band gap variation. The expansion of AB sites (increasing the energy favorable area) driven by $D$ field provides the driving force for decreasing $U$, since increasing AB site area can effectively lower $U$ to host two electrons with opposite spins in one AB site. In this scenario, shrinking the sizes of AB sites should cost extra energy.

### E. Transport measurement and carrier density estimation

For transport measurement, an AC excitation is set within the range from 0.2 mV to 5 mV at a frequency of 4.579 Hz. The current signal is probed by a lock-in amplifier. The voltage signals are detected through a low noise preamplifier SR550. The carrier density (if there is no internal polarization field in the sample) can be estimated by $\boldsymbol{n} = (C_{BG}V_{BG} + C_{TG}V_{TG})/e$, and the perpendicular electric field is expressed by $\boldsymbol{D} = (C_{BG}V_{BG} - C_{TG}V_{TG})/2\varepsilon_0 = D_B - D_T$. ($C_{TG}/C_{BG}$, top/bottom gate capacitance; $V_{TG}/V_{BG}$, top/bottom gate voltages). The top gate electric field is applied through the top few-layer hBN (defined by $D_T = C_{TG}V_{TG}/2\varepsilon_0$), and the back gate electric field is applied through the 300nm-SiO$_2$ layer on the Si substrate (defined by $D_B = C_{BG}V_{BG}/2\varepsilon_0$).

The dual-gate structure allows us to tune the charge density and the out-of-plane electric displacement field [S5]. Different from the multilayer graphene sensing scheme for detecting the interfacial ferroelectricity [S5, S6], the out-of-plane electrical polarization and/or potential difference between opposite polarization of domains in twisted TMDCs can be translated into the gate-controlled doping effect which is directly reflected by the electron transport hysteresis of the twisted TMDCs field-effect transistor devices [S7]. In our work, we measure electrical resistance at different carrier concentration under different DC gate scanning. The out-of-plane electronic polarization effects in our devices is presented by resistance hysteresis loops. For pristine few layer WSe$_2$ samples, no resistance hysteresis loop has been observed during forward-backward gate scanning (see FIG. S4). The moiré periodicity can be described by $\lambda = a/(\frac{2\sin\theta}{2})$, where $a$ is the lattice constant of WSe$_2$. The full filling carrier density is related to the moiré wavelength with $n_0 = 2/\frac{\sqrt{3}}{2}\lambda^2$. Then, we can further calibrate the twisted angle based on the estimation of the carrier densities.

### F. Interface and charge impurity effects

We have carried out different experiments to rule out the possibility that hBN interfaces or charge trapping effects could potentially generate a similar hysteretic resistivity as what we reported. Charge trapping effects are normally resulted from interface impurities. The interfaces in our devices are formed between hBN and WSe$_2$. We provide here more experimental data and analyses.

1. <u>The hBN moiré interface effects</u>: The hBN and WSe$_2$ layers have no specific twist angles in our devices. Therefore, there is no comparable moiré lattice formed at the hBN/WSe$_2$ interfaces. In fact, the lattice parameters of WSe$_2$ ($a = b = 0.3297$ nm) and hBN ($a = b = 0.2502$ nm) are very different. There is no comparable moiré superlattice formed in our devices.

2. <u>hBN/WSe$_2$/hBN interfaces:</u> We fabricated hBN-sandwiched WSe$_2$ field-effect devices using the same interface structure and measured the gate dependent transport properties. There is no hysteretic resistivity characteristics in these kind of samples during scanning the gate voltage (FIG. S4). Therefore, we confirmed that hBN/WSe$_2$ interfaces and the metal lead interfaces in our devices do not generate any hysteretic resistivity.

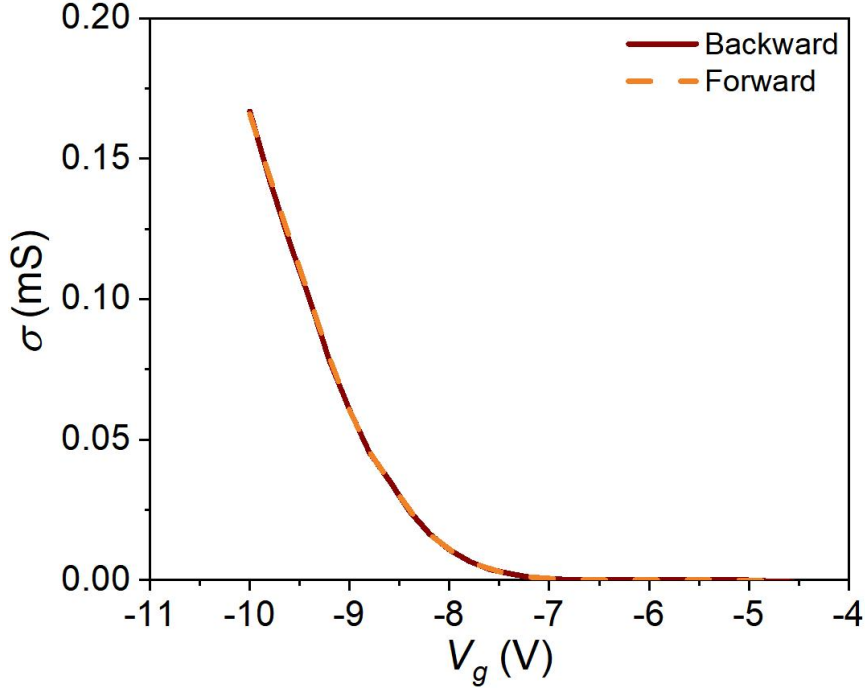

FIG. S4. Transport property of the reference hBN-sandwiched p-type WSe$_2$ field-effect device built based on the same interface structure used for twisted WSe$_2$. There is no hysteretic resistivity characteristic as measured by scanning the gate voltage at 1.6 K.

3. Charge trapping effects can normally be distinguished by changing the gate scanning rates. This method has been widely used in identifying charge trapping effects in graphene devices [S9]. We performed different gate scanning rates in our twist WSe$_2$ devices and did not observe any obvious difference as shown in FIG. S5. This is a strong evidence that charge trapping effects did not play a role in the observed hysteretic resistivity at half filling states.

4. Charge trapping can normally exist at room temperature such as in graphene [S9], in particular near the Dirac point. In this case, the charge trapping behavior becomes obvious. Near the resistance peak of a graphene device, charge traps cause shifts of the peak position (left-or-right). The amplitude of the peak has almost no change. This is because of the charge nature of the impurity at the device interfaces. However, in our device, the main feature of the hysteresis and resistance peaks is the amplitude change (high-or-low), a very different behavior.

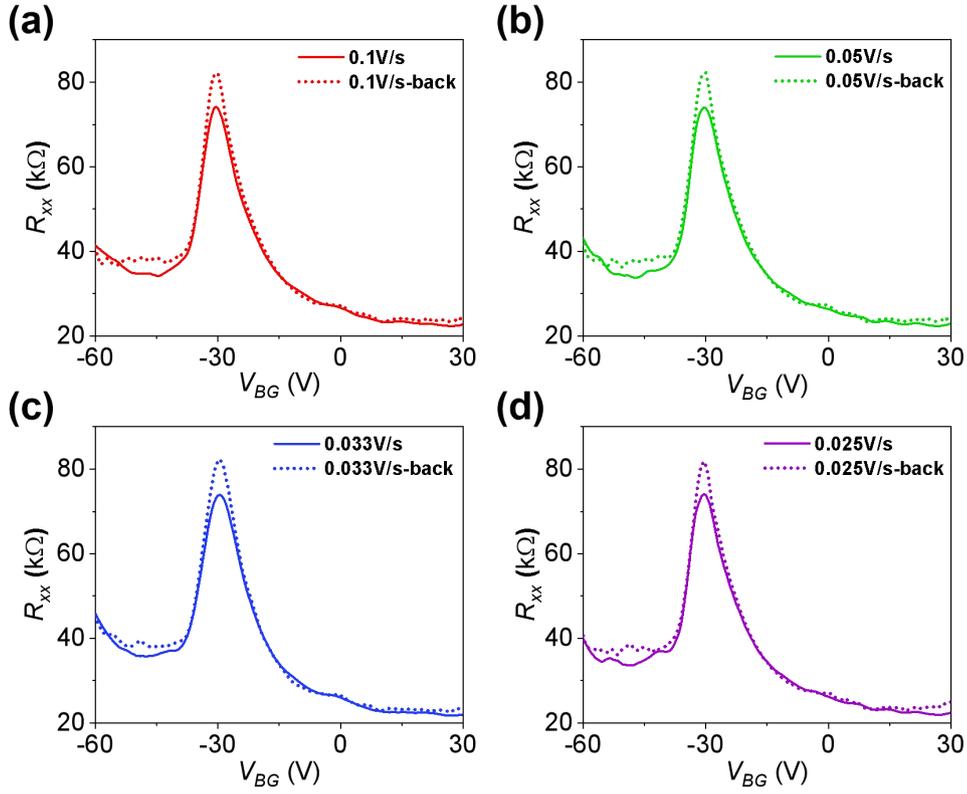

FIG. S5. *Transport measurements at different gate scanning rates in a 4.2° twist WSe$_2$ device. The half filling peaks and the hysteretic resistivity data do not show any obvious difference at different scanning rates.*

The hysteretic effect in our sample behaves differently compared to the normal charge trapping effects since it decays quickly when sample temperature is increased to about 40 K. However, the charge trapping effects caused by impurities normally can still exist at room temperature. In fact, charge trap effects usually occur globally in a device (not just at certain carrier concentration) in poor conducting materials, while our devices show high field-effect mobility with a high current injection efficiency between the electrodes and the twist bilayer WSe$_2$. The reproducibility of our device performance suggests that the hysteretic effects in our samples are not due to charge trapping.

### G. More transport results

Although the target of the twist angles is set to 4°, the fabricated devices often have a deviation from this angle. The twist angles of the antiparallel stacking moiré super-lattices we obtained range from 3.8 to 4.2°. Here are more experimental data we obtained from different devices. In FIG. S6(a), we show the half filling peaks measured at different gate voltage from a device with a 3.9° twist angle. FIG. S6(b) shows the corresponding resistance hysteresis. FIG. S7 shows the transport data taken from a device with a 4.2° twist angle. We find that for the samples with relatively large twist angles, their half filling peaks look sharp and the resistance hysteresis characteristics are more localized around the half filling density position. To calculate the displacement field $D$, we need to first estimate the top and bottom capacitances. By changing $V_{BG}$ and $V_{TG}$, we determine the $D$ filed by:

$$D = \frac{C_{BG}V_{BG} - C_{TG}V_{TG}}{2\varepsilon_0}$$

One of the results is shown in FIG. S8.

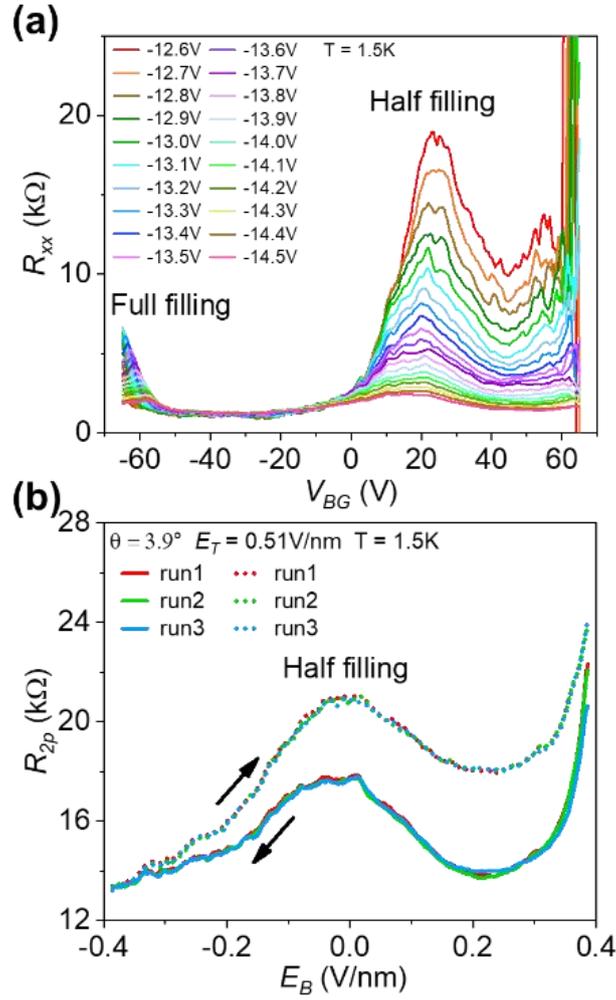

FIG. S6. (a) The half filling peaks measured at different displacement fields from a device with a 3.9° twist angle. (b) The resistance hysteresis near the half filling states.

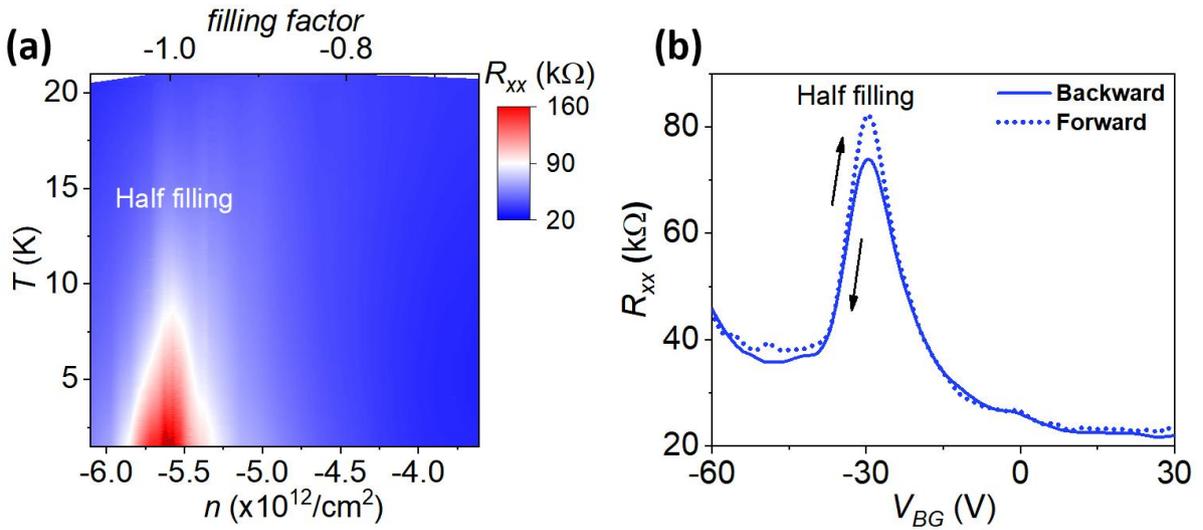

FIG. S7. Transport data taken from a 4.2° twist angle device. (a) Temperature effects on the half filling states. (b) The resistance hysteresis near the half filling states.

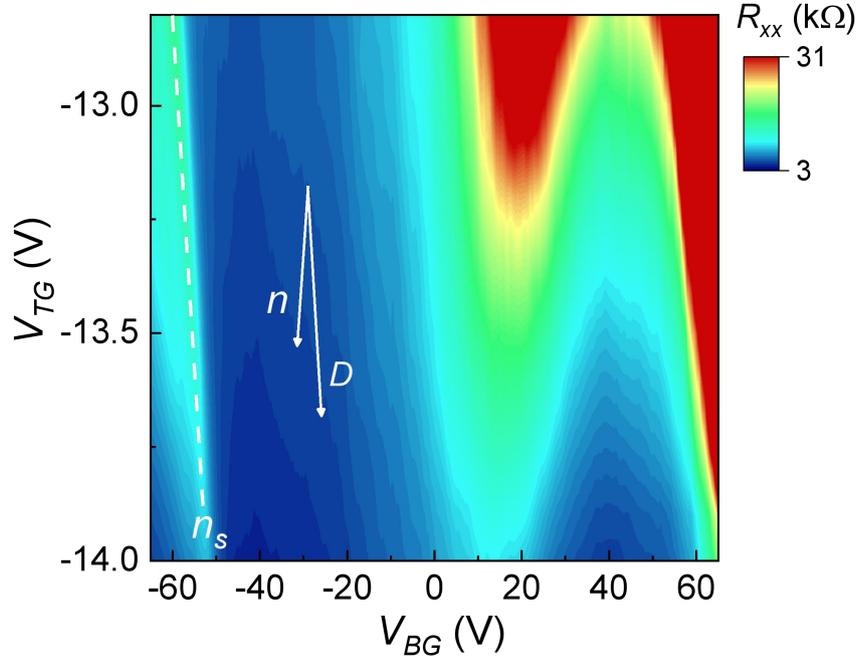

FIG. S8. Demonstration of both the strength of **D** field and the carrier density n dependent $R_{xx}$.

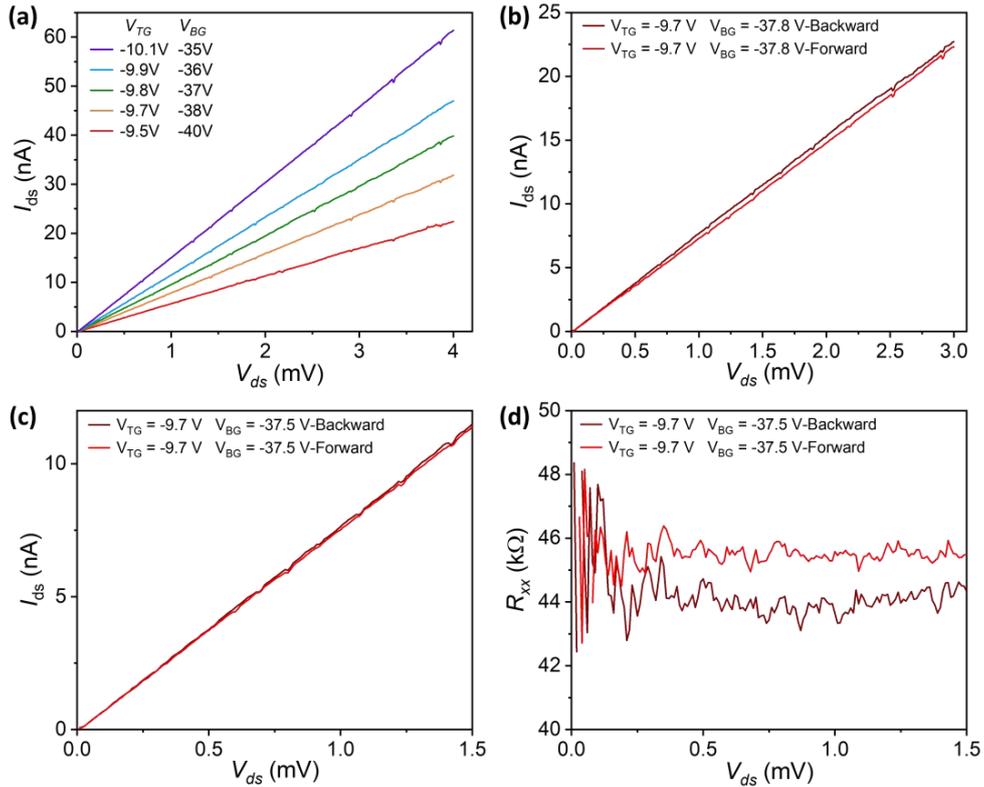

FIG. S9. I-V characteristics at the half filling states. (a) Transport measurements of I-V characteristics performed at the half filling states under the fixed top and back gate voltages. (b-c) I-V measurements of forward/backward scanning at ferroelectric state under different gate voltages. (d) Different resistances observed from forward/backward scanning.

We performed more transport measurements of I-V characteristics in the half filling states at fixed top gate (-9.5 V to -10.2 V) and back gate (-35 V to -41 V) voltages. We observed that for $V_{ds}$ larger than 0.01 mV, the I-V shows linear relationship. We also carried out I-V measurements on forward/backward scanning at the half filling ferroelectric state and observed different resistances.

In this study, we are not able to demonstrate a complete map of the resistant hysteresis loops as a function of ***D*** field because the top gate voltage has a very limited variation range. The main reason is that our devices are built with top and bottom gates. However, because there are totally four layers WSe$_2$, and the metal electrodes are designed to contact with the bottom layer of WSe$_2$, we need to apply a relative large top gate voltage in order to achieve a good ohmic contact to the semiconductor twist WSe$_2$. The ohmic contact behavior can only be kept in a narrow range of the top gate voltage (-10 V to -15 V). Top gate voltages larger than -15 V may cause breakdown of the dielectric layer. A top gate voltage smaller than -10 V resulting in poor electrical contact. Therefore, the scan including both top and bottom gate is mainly limited by the top gate. Scanning top gate in a large range will change the electrical contact characteristics and the experimental data are not reliable. Although we cannot provide a complete map of the resistant hysteresis loops as a function of ***D*** field, we add experimental data to partially show the resistant hysteresis changes at limited range of the ***D*** field. FIG. S10 shows the changes of resistance around half filling as a function of ***D*** field.

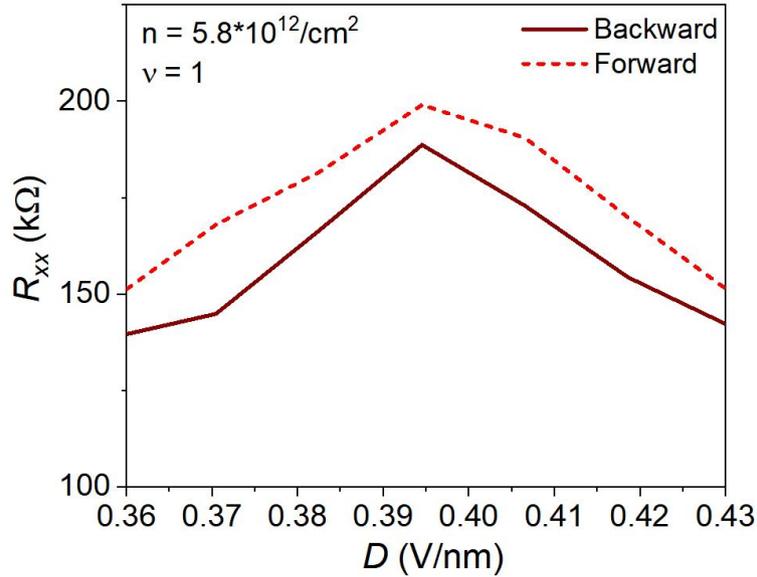

FIG. S10. *Experimental data showing the changes of resistance around half filling as a function of **D** field.*

### H. Interlayer charge transfer mechanism

FIG. S11(a) schematically shows the two bands associated with the bottom WSe$_2$ layer (directly connected to the electrodes) and the moiré super-lattice layer without considering interaction effects. Both bands are from the $\Gamma$ valley of WSe$_2$ with p-type characteristics. They form an isotype junction [S12, 13] for charge transfer between the two layers during forward/backward scanning of $V_{BG}$.

Due to the strong interaction occurring in the moiré super-lattice layer, LHB and UHB formed (FIG. S11(b)). For backward scanning of $V_{BG}$ (starting from +60 V), holes start to transfer from B-band to LHB first. In this case, both bands are p-type. Charge transfer across the p-p isotype junction is less resistive. Noticed that when holes occupy more than half of the states in LHB, the bottom band edge of LHB is electron-like (n-type semiconductor). Such a band alignment between LHB and B-band generates a n-p junction (anisotype junction). Holes transfer from B-band (p-type) to the n-type moiré layer is less resistive (similar to the forward biasing of a n-p junction). However, the opposite direction of holes transferring across the n-p junction is restricted. Charge transfer between B-band to UHB is less resistive since their junction is always isotype.

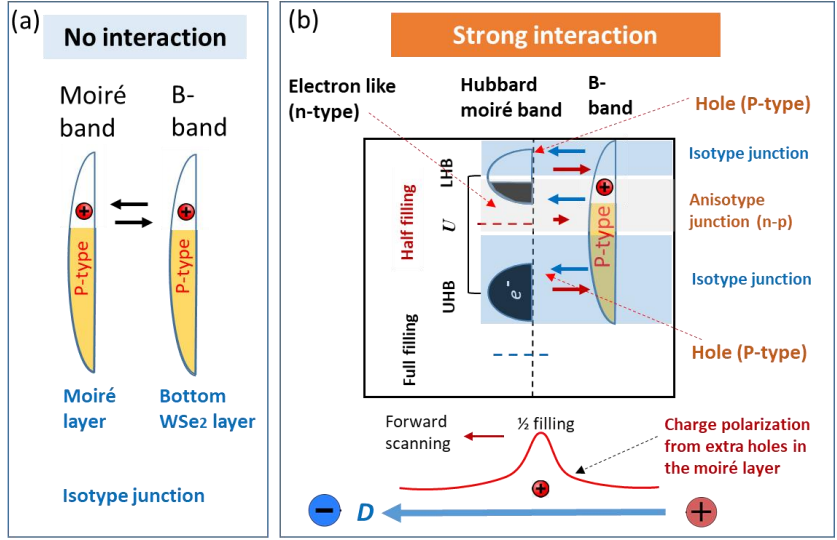

FIG. S11. (a) Two bands associated with the bottom WSe$_2$ layer and the moiré super-lattice without interaction. (b) Strong interaction induces LHB and UHB of the Hubbard bands. Charge transfers between the Hubbard bands and B-band near the half filling states is modulated by the formation of the Hubbard band gap.

For forward scanning of $V_{BG}$, holes transfer back to the B-layer. In this case, charge transfer across all isotype junctions is less resistive. However, near half filling states, the charge transfer is severely restricted (analogous to the reverse biasing of a n-p junction). The consequence is that there are more holes retained in the moiré layer (in the same time the holes in the B-layer are released to the electrodes normally) when a forward scanning approaches the half filling region. Therefore, the charge polarization between the moiré layer and the B-layer is established. The extra holes in the moiré layer countervail only a part of the strength of the **D** field (FIG. S11(b)), resulting in a higher $R_{xx}$ compared to that in the opposite direction of backward scanning $V_{BG}$. This is because $R_{xx}$ is inversely proportional to the **D** field as revealed experimentally in Fig. 5(c) in the main text.

We noticed that there is also resistance hysteresis in the metallic phase. First, bilayer WSe$_2$ is semiconductor. The so-called insulating states at half filling are generated by the correlation effects in the twist moiré system. The nearby metallic states are also called strongly correlated metallic phases, such as the superconductor states in twisted WSe$_2$ bilayer we discovered. Second, since the electron polarization states are relevant to the correlated states, it is reasonable to believe that the resistive hysteresis occurring at states away from the half filling states is also related to correlation effect. However, the resistive hysteresis occurring at the state far away from the half filling states is indeed complicated. According to our model, the alignment of the Hubbard band and the additional B-band in the "metallic states" is isotype junction. If there is a slight barrier between the moiré interface layers and the B-band layer which might be due to the formation of the moiré bands, the electron polarization could happen, resulting in the resistance hysteresis. This situation is similar to that of the anisotype junction.